\documentclass[prl,twocolumn,showpacs]{revtex4}

\usepackage{graphicx}% Include figure files
\usepackage{dcolumn}% Align table columns on decimal point
\usepackage{bm}% bold math

\begin{document}

\preprint{APS/123-QED}

\title{Quantum confinement corrections to the capacitance of
gated one-dimensional nanostructures}

\author{K. M. Indlekofer}
\email{m.indlekofer@fz-juelich.de}
\author{J. Knoch}
\affiliation{
Institute of Thin Films and Interfaces (ISG-1) and
Center of Nanoelectronic Systems for Information Technology (CNI),
Research Centre J\"ulich GmbH, D-52425 J\"ulich, Germany
}
\author{J. Appenzeller}
\affiliation{
IBM T. J. Watson Research Center, P.O. Box 218,
Yorktown Heights, New York 10598
}

\date{\today}

\begin{abstract}
With the help of a multi-configurational Green's function approach
we simulate single-electron Coulomb charging effects in gated
ultimately scaled nanostructures which are beyond the scope of a
selfconsistent mean-field description. From the simulated
Coulomb-blockade characteristics we derive effective system
capacitances and demonstrate how quantum confinement effects give
rise to corrections. Such deviations are crucial for the
interpretation of experimentally determined capacitances and the
extraction of application-relevant system parameters.
\end{abstract}

\maketitle

One of the major challenges for the simulation of electronic
transport in low-dimensional nanostructures consists in an adequate
description of the Coulomb interaction. In ultimately scaled
semiconductor nanosystems, only a few electrons constitute the
current. Thus, the details of the electron-electron interaction may
become crucial for the electronic properties, as can be seen in
experimentally observed single-electron charging effects in carbon
nanotubes and III/V nanowhiskers \cite{suz02,the03}. The channel
region of a typical application-relevant nanostructure involves on
the order of 100 single-particle states (or sites) and can be highly
inhomogeneous and anisotropic. Here, external contacts and gate
electrodes in general introduce non-linear perturbations to the
system. Hence, highly idealized interacting few-level models with a
small number of effective parameters (constants) become inadequate.
For a realistic simulation of quantum transport in
application-relevant nanodevices, various approaches have been used,
differing in the formulation of Coulomb interaction and contact
coupling. While the orthodox theory \cite{ben91,ave91} correctly
describes few-electron effects such as single-electron tunneling in
quasi-isolated quantum dot systems, the nonequilibrium Green's
function formalism (NEGF) \cite{lak97,datta,ven02,lak05} also
accounts for contact renormalization and dissipation terms. However,
the NEGF becomes unable to describe the details of charging effects
of a few fluctuating electrons as soon as a mean-field approximation
is introduced.
%In principle, a real-time renormalization group approach
%\cite{sch00} should be able to overcome these limitations.

\begin{figure}[ht]
\vspace{-0.0cm}
\begin{center}
\includegraphics[width=5.0cm]{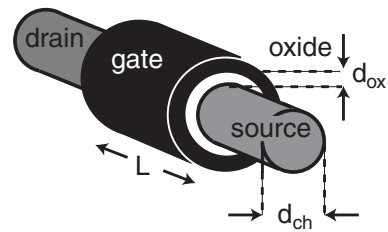}
\end{center}
\vspace{-0.3cm} \caption{\footnotesize \label{fig:system} Schematic
sketch of the considered coaxially gated  nanosystem (see text for
more details).
% The radial
%cut through the channel region visualizes the local density of
%states (LDOS) and the groundstate probability density. The solid
%white line corresponds to the mean-field potential. Here, the lowest
%LDOS resonance is occupied with one electron (which is not subject
%to its own Hartree potential).
}
\end{figure}
Recently, we have presented a multi-configurational approach (MCSCG)
\cite{ind05}, which represents a consistent extension of the
mean-field NEGF for the inclusion of single-electron effects under
application-relevant conditions. This approach combines the
simplicity and scalability of the mean-field NEGF with a many-body
Fock space description of the Coulomb interaction of those electrons
that are resonantly trapped within the nanostructure. As a key
element, the algorithm identifies ``relevant'' trapped
single-particle states that are subject to occupation fluctuations.
Within the resulting relevant Fock subspace, many-body
``configurations'' $\kappa$ and associated weights $w_{\kappa}$ are
defined as eigenstates and eigenvalues of the projected many-body
statistical operator $\rho^{MB}$, respectively. Consequently, the
system Green's functions are written as weighted averages:
$\bar{G}=\sum_{\kappa} w_{\kappa} G[\kappa]$, where $G[\kappa]$
follows from a Dyson equation with a suitable self-energy
approximation $\Sigma[\kappa]$. As a self-consistency condition,
$\rho^{MB}$ is obtained by use of the projected many-body
Hamiltonian $H^{MB}$ and the subsidiary condition that $G^{MB}$
should resemble $\bar{G}$ within the relevant subspace, where
$G^{MB}$ is calculated via $\rho^{MB}$ and $H^{MB}$.

In this article, we show how the MCSCG formalism can be used to
calculate electronic transport properties of gated one-dimensional
(1D) nanosystems in the single-electron tunneling regime beyond the
mean-field approach. Furthermore, we extract realistic system
capacitances from the simulated I-V characteristics. Three length
scales are relevant for the Coulomb interaction energies of the
system: The geometrical length $L$, the electrostatic screening
length $\lambda$, and the de Broglie wavelength of the electron. The
latter becomes important in a nanoscale system with quantum
confinement, giving rise to significant deviations from the expected
capacitances.

In the following, we discuss a 1D nanostructure (i.e., one {\it
lateral} mode only) as depicted in Fig.~\ref{fig:system}. In the
shown coaxial geometry, the channel is surrounded by a metallic gate
electrode, separated by an insulating oxide layer. The 1D channel of
the nanostructure is connected to two Schottky contacts (denoted as
source and drain) which serve as electron reservoirs \footnote{The
considerations remain valid qualitatively for general contact
barriers.}. In order to visualize the electronic structure of the
system, a radial cut through the inner region of the structure has
been inserted, showing the groundstate probability density
$|\psi(x)|^2$ and the associated local density of states (LDOS) for
1 electron inside the channel. For the discussion of interaction
effects we will now consider an example of an ultimately scaled
semiconducting nanocolumn (see also \cite{ind05}) with an effective
channel length of $L=$ 19nm, a channel diameter of $d_{ch}=$ 4nm and
an oxide thickness of $d_{ox}=$ 10nm. Furthermore, we chose
$\epsilon_{ch}=$ 15 and $\epsilon_{ox}=$ 3.9, yielding a screening
length of $\lambda\approx$ 3.7nm. The latter is obtained from
\begin{equation}
\label{eq:lambda}
\lambda^2=
\frac{\epsilon_{ch}}{\epsilon_{ox}}
\frac{d^2_{ch}}{8}
~\mbox{ln}\left(1+2\frac{d_{ox}}{d_{ch}}\right)
\end{equation}
for a coaxial system layout \cite{aut97}. The Schottky source and
drain contacts exhibit a barrier of $\Phi_B=$ 0.5eV (with respect to
the Fermi energy), whereas the effective electron mass is taken as
$m^*=0.05 m_e$. This parameter set was chosen in order to clearly
visualize the Coulomb effects, at the same time being close enough
to experimental realizations.

\begin{figure}[ht]
\vspace{-0.0cm}
\begin{center}
\includegraphics[width=8.5cm]{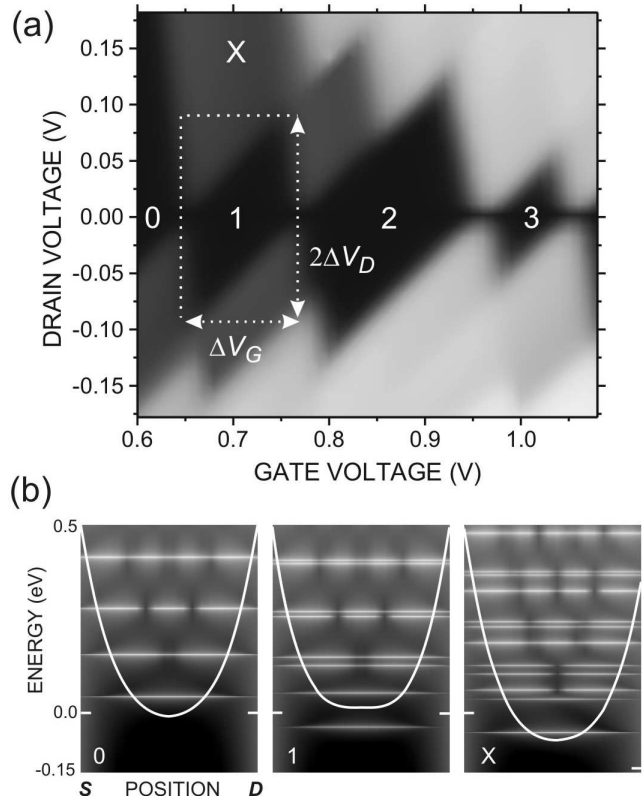}
\end{center}
\vspace{-0.2cm} \caption{\footnotesize \label{fig:couldiamonds} (a)
Simulated single-electron tunneling characteristics as a function of
the applied gate and drain voltages for $T=$77K. Here, the absolute
value of the drain current is mapped to a non-linear grayscale.
Within the black diamonds of Coulomb blockade the electron number
becomes quantized as indicated by white numbers. $\Delta V_G$ and
$\Delta V_D$ denote the width and half height of the first diamond,
respectively. (b) Corresponding local density of states and
mean-field potential within the channel for selected points 0,1,X in
map (a). Point X corresponds to a nonequilibrium state with electron
number $\approx$ 0.3.}
\end{figure}
Fig.~\ref{fig:couldiamonds}(a) shows a simulated grayscale map of
the drain current as a function of the applied gate and drain
voltages in the single-electron tunneling regime ($T=$ 77K)
\footnote{In this example, up to 64 configurations of relevant
states with a diagonal interaction approximation were considered.}.
One can clearly identify the typical diamond-like shaped black areas
of Coulomb blockade with a suppressed drain current. Dark-gray areas
correspond to electronic transport through the lowest longitudinal
resonantly-trapped state, whereas brighter areas involve the first
(and higher) excited resonances. Such a behavior is beyond the
capabilities of a mean-field NEGF that is based on a selfconsistent
potential, demonstrating the strengths of the employed
multi-configurational approach. In order to visualize the
interaction mechanism that goes beyond mean-field,
Fig.~\ref{fig:couldiamonds}(b) shows the local density of states
(LDOS) within the channel region at different voltage conditions and
the associated mean-field potentials. For the empty channel "0", we
obtain a set of quasi-bound states and resulting resonances. With
one electron inside the channel near equilibrium point "1", one
electron occupies the lowest state and all further states are
shifted to higher energies by the Coulomb interaction. Note that the
electron does not experience its own Hartree potential. Finally,
under strong non-equilibrium bias conditions at point "X", the
spectral properties result from a weighted mixture of an empty
channel and a channel containing one electron. Here, the electronic
properties clearly deviate from a mean-field picture.

For the considered example, the height of the Coulomb diamond with 1
electron is determined by the Coulomb interaction of two electrons
with opposite spin within the lowest channel resonance
(groundstate). In contrast, for the second diamond one has to
consider not only the interaction between the lowest and the first
excited states but also the difference between the single-particle
energies of the two involved states. Hence, its size is
significantly enlarged.
%For 3 electrons, however, the Coulomb
%interaction of two electrons with opposite spin in the first excited
%state has to be taken into account. Due to the increased spatial
%extent of the wave function of the excited state, the size of the
%Coulomb diamond for 3 electrons is reduced compared to the case of 1
%electron inside the channel.
These observations strongly depend on the degeneracy of the
considered channel resonances and on the ratio of quantization and
Coulomb energies. In ultimately scaled systems that are addressed in
this letter, the longitudinal quantization energy dominates and it
is justified to consider the groundstate Coulomb matrix element
only. In contrast, for longer channels, the quantization energy may
become much smaller than the charging energy and we could end up
with a situation where the occupation of higher excited states (with
a reduced Coulomb matrix element) and a re-organization of electrons
might become energetically favorable for the addition of the next
electron \cite{ind00} (and Wigner molecules might form
\cite{jar04}).
%In such a case, the Coulomb diamonds typically
%become equal in size for higher occupation (and Wigner molecules
%might form \cite{jar04}).
In the following, we restrict ourselves to the discussion of the
lowest Coulomb diamond which corresponds to the onset of the current
flow.

In general, effective quantities such as the gate capacitance
$C^*_G$ and the total system capacitance $C^*_{\Sigma}$ can be
extracted from the shape of the observed Coulomb diamonds
\cite{grabert} (provided that the latter is solely due to Coulomb
interaction): From the total width $\Delta V_G$ and {\it half}
height $\Delta V_D$ one obtains $C^*_G=e/\Delta V_G$ and
$C^*_{\Sigma}=e/\Delta V_D$, respectively. In our example
(Fig.~\ref{fig:couldiamonds}), we have $\Delta V_G\approx$ 114mV and
$2\Delta V_D\approx$ 186mV, and therefore $C^*_G\approx$ 1.41aF and
$C^*_{\Sigma}\approx$ 1.72aF. Consequently, a Coulomb interaction
energy of $E^*_{\Sigma}=e^2/C^*_{\Sigma}\approx$ 93meV can be
derived (which is identical to a two-particle Coulomb
matrix-element, see Ref.~\cite{ind05}). Note that for such
ultra-short nanosystems, where the quantization effects play an
important role, these capacitances in general are voltage dependent
quantities, depending on the wave functions of the involved
longitudinal channel resonances that are being filled at the
considered electrode voltages.

For the classical capacitance of the considered coaxial geometry
of the 1D channel region one obtains
\begin{equation}
\label{eq:cox}
C_{ox}=\frac{\pi\epsilon_0\epsilon_{ch}d^2_{ch}L}{4\lambda^2}
=
2\pi\epsilon_0\epsilon_{ox}
\frac{L}{\mbox{ln}\left(1+2\frac{d_{ox}}{d_{ch}}\right)}.
\end{equation}
Employing the Coulomb Green's function (for details see Ref.~\cite{ind05})
for fixed potential boundary conditions,
the Coulomb interaction energy $E^{hom}_{\Sigma}$
of a homogeneously distributed
test charge $e$ with an identical background charge thus reads as
$E^{hom}_{\Sigma}=e^2/C^{hom}_{\Sigma}$ with
the total homogenous system capacitance
\begin{equation}
\label{eq:chom}
C^{hom}_{\Sigma}=C_{ox}
\left(1-2\frac{\cosh(\frac{L}{\lambda})-1}
{\frac{L}{\lambda}\sinh(\frac{L}{\lambda})}\right)^{-1}.
\end{equation}
(Note that $C^{hom}_{\Sigma}>C_{ox}$
and $C^{hom}_{\Sigma}\to C_{ox}$ for $L/\lambda\to\infty$.)
Using the given parameters, we thus obtain $C_{ox}=$ 2.30aF and
$C^{hom}_{\Sigma}=$ 3.75 aF.
In this context, we therefore define the quantity
\begin{equation}
\label{eq:qcf}
f_{\Sigma}=\frac{e^2}{C^*_{\Sigma}}/\frac{e^2}{C^{hom}_{\Sigma}}
=C^{hom}_{\Sigma}\Delta V_D/e,
\end{equation}
which we refer to as the ``quantum confinement factor'' for the
given geometrical structure and parameters (voltages, etc.). In the
example above, we have $f_{\Sigma}\approx$ 2.2. The observation that
$f_{\Sigma}$ becomes $>1$ for ultimately scaled structures can
easily be understood if one takes a look at the quasi-bound
single-electron wave function (Fig.~\ref{fig:couldiamonds}(b)) of
the considered Coulomb diamond: Typically, this wave function
exhibits an enhanced probability amplitude in the center region of
the 1D channel due to quantum confinement in transport direction
which is a result of the Schottky-barriers at the source and drain
contacts. In other words, the electron is spatially confined within
a reduced effective channel length $<L$. This implies an increased
Coulomb interaction energy $E^*_{\Sigma}=$ 93meV compared to
$E^{hom}_{\Sigma}=$ 43meV for a homogenous charge distribution, or
equivalently $C^*_{\Sigma}<C^{hom}_{\Sigma}$ \footnote{The reduction
of $C^*_{\Sigma}$ with respect to $C^{hom}_{\Sigma}$ is not to be
confused with the concept of a finite quantum capacitance $C_Q=e^2
\bar{D}(E_F)$ with a thermally averaged density of states
$\bar{D}(E_F)$ \cite{lur87}. The considered nanosystem in the
single-electron regime exhibits a quasi-zero-dimensional electronic
structure, where $C_Q$ equals 0 or $\infty$. }. Consequently, larger
voltage changes are required to add one further electron to the
channel region. This implies that care has to be taken with the
extraction of geometrical parameters from experimental data as will
be discussed in more detail below.

\begin{figure}[ht]
\vspace{-0.0cm}
\begin{center}
\includegraphics[width=8.0cm]{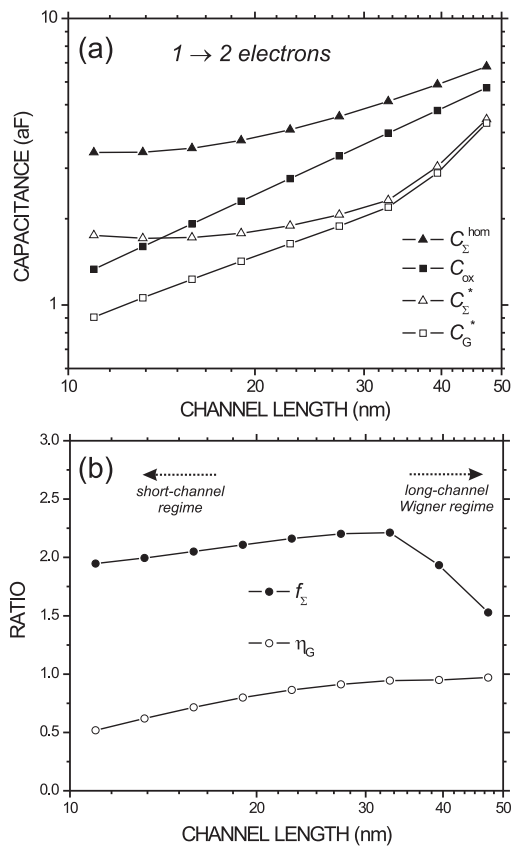}
\end{center}
\vspace{-1.0cm} \caption{\footnotesize \label{fig:cap} (a) Simulated
capacitances for the first two electrons as a function of the
channel length $L$ with an adaptive gate voltage. The screening
length $\lambda\approx$ 3.7nm and the effective mass $m^*=$
0.05$m_e$ are kept constant. (b) Quantum confinement factor
$f_{\Sigma}=C^{hom}_{\Sigma}/C^*_{\Sigma}$ and gate efficiency
$\eta_G=C^*_{G}/C^*_{\Sigma}$ as derived from (a). }
\end{figure}
Fig.~\ref{fig:cap}(a) shows the simulated capacitances for the first
two electrons (groundstate wave function and next higher state) as a
function of the channel length $L$ for a fixed $\lambda$ and $m^*$.
Here, the gate voltage $V_G$ is chosen adaptively such that the
groundstate energy coincides with the Fermi energy of the source
contact (for a constant Schottky barrier height $\Phi_B=$ 0.5eV with
$V_D\equiv V_S\equiv 0$). For each $L$, the wave function and the
associated Coulomb interaction energy
$E^*_{\Sigma}=e^2/C^*_{\Sigma}$ have been obtained via numerical
diagonalization of the channel Hamiltonian \cite{ind05}. As
expected, for larger $L/\lambda$ we observe that $C^*_{\Sigma}\to
C^*_{G}$ and $C^{hom}_{\Sigma}\to C_{ox}$, which means that the
source and drain influence becomes negligible in this limit. On the
other hand, for the ultra-short structure with $L<$ 20nm, one can
clearly identify the increasing influence of the source and drain
capacitances. Therefore, an increased gate voltage becomes necessary
to pull the lowest quasi-bound state down to the Fermi energy of the
source contact for a fixed Schottky barrier height (e.g., $V_G=$
0.52V for $L=$ 50nm, and $V_G=$ 1.1V for $L=$ 10nm). This is
described quantitatively by the efficiency
$\eta_G=C^*_G/C^*_{\Sigma}=\Delta V_D/\Delta V_G$ of the gate
electrode at changing the considered quasi-bound energy level, as
plotted in Fig.~\ref{fig:cap}(b).

Most noticeably for all ranges of $L/\lambda$, there is a
significant deviation of $C^*_{\Sigma}$ from $C^{hom}_{\Sigma}$ (and
$C^*_G$ from $C_{ox}$) as can be seen in Fig.~\ref{fig:cap}(a). This
observation, in fact, is described by the quantum confinement factor
$f_{\Sigma}$ as plotted in Fig.~\ref{fig:cap}(b).
%A rough estimation
%of $f_{\Sigma}$ can be obtained from an idealized hard-wall quantum
%well wave function of the form $\psi_{box}(x)=\sqrt{2/L}\sin(\pi
%x/L)$. By using the Coulomb Green's function of Ref.~\cite{ind05},
%we obtain $f_{\Sigma}\to 3/2$.
In the short-channel regime, we obtain $f_{\Sigma}>1$, which can be
explained by the longitudinal confinement of the groundstate wave
function and the resulting increased Coulomb interaction.
%in comparison to $\psi_{box}$ (stemming from the exponential tails of
%the Schottky barriers at the source and drain contacts).
Furthermore, as can be seen in Fig.~\ref{fig:cap}, the onset of the
short-channel regime, where $\eta_G$ deviates significantly from 1
due to increasing source and drain influence, is shifted to smaller
channel lengths $L<$ 30nm compared to $L>$ 50nm where
$C^{hom}_{\Sigma}$ starts to deviate from $C_{ox}$. Again, this
finding can be attributed to the longitudinal confinement of the
wave function with a reduced electron charge density at the source
and drain contacts. For larger channel lengths $L$, however, we
observe a rapid decrease of $f_{\Sigma}$ (or equivalently, a strong
increase of $C^*_{\Sigma}$ and $C^*_G$). This finding can be
attributed to the onset of a Wigner regime due to a significantly
reduced average electron density (because of an increasing spatial
separation of the two considered electrons compared to $\lambda$).

The discussed confinement effect becomes very important for the
experimental determination of application-relevant capacitances from
measured Coulomb diamonds. Analogous to the interpretation of our
simulation results above, effective capacitances $C^*_{\Sigma}$ and
$C^*_G$ can be calculated from the experimentally determined values
of $\Delta V_D$ and $\Delta V_G$ at low temperatures (typically
$4k_BT<E^*_{\Sigma}$). One has to be careful with the interpretation
of these effective capacitances as far as the derivation of
geometrical parameters from the latter and their relevance at
room-temperature are concerned. For example, it is possible to
calculate $d_{ox}$ from $C^*_{\Sigma}$. By inspecting
Eqs.~(\ref{eq:lambda}-\ref{eq:qcf}), however, one can see that this
is a non-trivial task since $\lambda$ depends on $d_{ox}$, and
$f_{\Sigma}$ (which results from a Coulomb matrix-element) in turn
depends on $\lambda$. If we identified $C^{hom}_{\Sigma}$ with
$C^*_{\Sigma}\equiv e/\Delta V_D$ (i.e., setting $f_{\Sigma}\equiv
1$) from experimental data, we would overestimate $d_{ox}$. In the
considered example, we would end up with $d_{ox}>$ 50nm (instead of
10nm) which is a consequence of the logarithm in
Eqs.~(\ref{eq:lambda},\ref{eq:cox}).
%As a second example, for a
%transistor application of the considered nanostructure one typically
%needs to know the gate capacitance. However, $C^*_G$ determined from
%the low-temperature single-electron characteristics need not be the
%same as the gate capacitance $C_G=\partial Q/\partial V_G$ at room
%temperature. For $4k_BT\gtrsim E^*_{\Sigma}$ one typically has to
%deal with a semi-classical transport situation, where the particular
%wave function that was considered for the determination of a single
%$C^*_G$ might become irrelevant. Here, $C_G$ is the result of
%multiple thermally averaged states with different Coulomb
%interaction energies. In particular, $C_G$ is reduced if the Fermi
%energy approaches the lowest quasi-bound state (lower energy
%cutoff). This effect is due to a reduction of available states
%within a $4k_B T$ window at $E_F$.
%For the discussed example we get $C_G\approx$ 1.1aF (calculated for
%$V_G=$ 0.71V at $T=$ 300K) instead of $C^*_G\approx$ 1.4aF of the
%first Coulomb diamond.

In conclusion, we have demonstrated that the multi-configurational
selfconsistent Green's function approach (MCSCG) is able to model
single-electron Coulomb charging effects in ultimetely scaled gated
1D nanostructures, providing a means for the determination of
interaction energies and resulting effective capacitances. In turn,
a quantitative discussion of quantum confinement corrections to the
system capacitance was given, resulting in a significant deviation
from the classical capacitance which has an immediate consequence
for the interpretation of experimental system parameters.

%\bibliography{paper2}

\end{document}